\documentclass[aps, prc, superscriptaddress, twocolumn, floatfix, showpacs, longbibliography, altaffilletter, groupedaddress]{revtex4-1}
\usepackage[american]{babel}
\usepackage[T1]{fontenc}
\usepackage[utf8]{inputenc}
\usepackage{textcomp}

\usepackage{amsmath} 
\usepackage{amssymb}
\usepackage{braket}

\usepackage{blindtext}

\usepackage{isotope} 
\usepackage{bm} 
\usepackage{natbib}
\usepackage[squaren]{SIunits} 
\usepackage{booktabs} 
\usepackage{graphicx}
\usepackage{subfigure}
\usepackage[ulem=normalem]{changes}
\usepackage[percent]{overpic}
\usepackage{wasysym}  

\usepackage{scrextend}  

\usepackage{tabularx}
\usepackage{longtable}
\usepackage{multirow}
\usepackage{dcolumn}
\usepackage{color} 
\usepackage[colorlinks,hyperindex]{hyperref} 
\hypersetup
    {
        colorlinks,%
        citecolor=blue,%
        linkcolor=blue,%
        urlcolor=blue,%
    }

\newcommand{\be}[4]{B(E2;#1^+_#2 \to #3^+_#4) }

\begin{document}

\title{Tests of collectivity in $^{98}$Zr by absolute transition rates}
\date{\today}

\author{V.~Karayonchev}

    \affiliation{Institut f{\"u}r Kernphysik, Universit{\"a}t zu K{\"o}ln, 50937 K{\"o}ln, Germany}     
\author{A.~Leviatan}
\author{N.~Gavrielov}
\affiliation{Racah Institute of Physics, The Hebrew University, 
Jerusalem 91904, Israel}     
      
\author{J.~Jolie}
    \affiliation{Institut f{\"u}r Kernphysik, Universit{\"a}t zu K{\"o}ln, 50937 K{\"o}ln, Germany}  
\author{A.~Blazhev}
    \affiliation{Institut f{\"u}r Kernphysik, Universit{\"a}t zu K{\"o}ln, 50937 K{\"o}ln, Germany} 

   \author{A.~Dewald}
    \affiliation{Institut f{\"u}r Kernphysik, Universit{\"a}t zu K{\"o}ln, 50937 K{\"o}ln, Germany}   

\author{A.~Esmaylzadeh}
    \affiliation{Institut f{\"u}r Kernphysik, Universit{\"a}t zu K{\"o}ln, 50937 K{\"o}ln, Germany}    
 
\author{C.~Fransen}
    \affiliation{Institut f{\"u}r Kernphysik, Universit{\"a}t zu K{\"o}ln, 50937 K{\"o}ln, Germany}     
       
\author{G.~Häfner}
    \affiliation{Institut f{\"u}r Kernphysik, Universit{\"a}t zu K{\"o}ln, 50937 K{\"o}ln, Germany}       
 
\author{L.~Knafla}
    \affiliation{Institut f{\"u}r Kernphysik, Universit{\"a}t zu K{\"o}ln, 50937 K{\"o}ln, Germany}       
       
\author{J.~Litzinger }
    \affiliation{Institut f{\"u}r Kernphysik, Universit{\"a}t zu K{\"o}ln, 50937 K{\"o}ln, Germany}   
       
\author{C.~M\"uller-Gatermann}
    \altaffiliation[Present address: ]{Physics Division, Argonne National Laboratory, Argonne, Illinois 60439, USA}
    
        \author{J.-M.~R\'egis}
    \affiliation{Institut f{\"u}r Kernphysik, Universit{\"a}t zu K{\"o}ln, 50937 K{\"o}ln, Germany}
        
\author{K.~Schomacker}
    \affiliation{Institut f{\"u}r Kernphysik, Universit{\"a}t zu K{\"o}ln, 50937 K{\"o}ln, Germany}
         
\author{A.~Vogt}
    \affiliation{Institut f{\"u}r Kernphysik, Universit{\"a}t zu K{\"o}ln, 50937 K{\"o}ln, Germany}        
        
\author{N.~Warr}
    \affiliation{Institut f{\"u}r Kernphysik, Universit{\"a}t zu K{\"o}ln, 50937 K{\"o}ln, Germany}

\begin{abstract}
Lifetimes of low-spin excited states in $^{98}$Zr were measured
using the recoil-distance Doppler-shift technique and the Doppler-shift attenuation method.
The nucleus of interest was populated
in a $^{96}$Zr($^{18}$O,$^{16}$O)$^{98}$Zr two-neutron transfer reaction
at the Cologne FN Tandem accelerator. Lifetimes of six low-spin excited states, of which four are unknown, were measured. The deduced $B(E2)$ values were compared with Monte Carlo shell model and interacting boson model with configuration mixing calculations. Both approaches reproduce well most of the data but leave challenging questions regarding the structure of some low lying states. 
\end{abstract}

\maketitle

\section{Introduction}
\begin{figure}[h!]
\includegraphics[width=\linewidth]{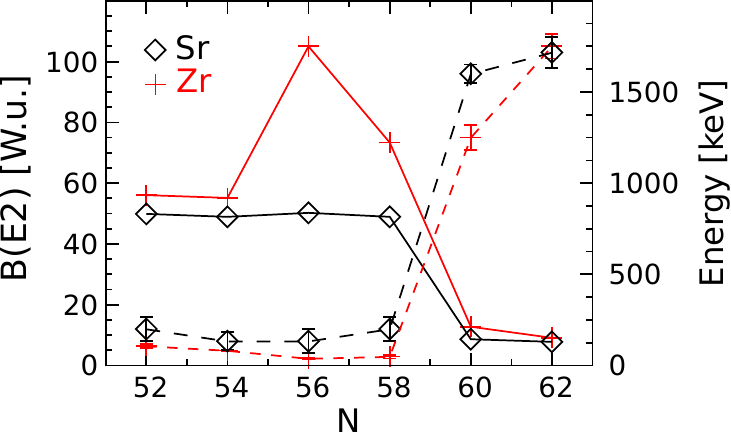}
\caption{The energies of the first excited 2$^+$ states for Zr and Sr isotopes with $N$ = 52$-$62 (symbols connected by a solid line), together with the $B(E2;2^+_1\rightarrow0^+_1)$ (symbols connected by a dashed line). Data are taken from the nuclear data sheets~\cite{NuDat_A90,NuDat_A92,NuDat_A94,NuDat_A96,NuDat_A98,NuDat_A100,NuDat_A102}. The $B(E2;2^+_1\rightarrow0^+_1)$ value for $^{98}$Zr is taken from~\cite{Singh2018}.}
\label{fig:BE2s}
\end{figure}

Quantum shape-phase transition is a phenomenon present in many physical systems including the atomic nucleus~\cite{Cejnar2010,Heyde2011}. Depending on the proton and the neutron numbers, the ground state of the nucleus can have different shapes. Nuclei having neutron or proton number close to the magic numbers tend to exhibit a spherical ground state. As one moves away from a closed shell, towards mid-shell, the number of available states to mix under the residual interaction grows rapidly and collectivity starts to develop. At this point, the proton-neutron correlations start to become dominant, making a deformed shape energetically more favorable and the ground state becomes deformed. Although the development of collectivity is usually a gradual process, the Zr and Sr isotopes are unique on the nuclear chart as they experience a very rapid onset of collectivity when crossing neutron number $N$ = 60. This is well observed through the systematics of the $B(E2;2^+_1\rightarrow0^+_1)$ values (see Fig.~\ref{fig:BE2s}). Starting from the shell-closure at $N$ = 50 until $N$ = 58, both the Zr and the Sr isotopes have rather low transition probabilities and correspondingly high excitation energies, consistent with a spherical configuration. With the addition of only two neutrons beyond $N$ = 58 the $B(E2;2^+_1\rightarrow0^+_1)$ values for both isotopes jump abruptly to the collective values of about 100 W.u., consistent with a deformed ground-state configuration. The increase of the transition probabilities is accompanied by a sharp decrease of the excitation energies of the $2^+_1$ to values typical for a rotational nucleus of the mass region. Phenomenologically, this could be interpreted as a coexistence of nuclear configurations with different shapes. For $N$ < 60  the ground state is spherical and a deformed configuration has higher energy. This deformed configuration lowers in energy as neutrons are added and eventually becomes the ground state, while the spherical configuration is pushed higher in energy. Indeed, low-lying excited $0^+$ states are observed in the region and their energy drops sharply as $N$ = 60 is approached. The scenario of having an excited deformed configuration was recently confirmed by an electron-scattering experiment on the neighboring nucleus $^{96}$Zr~\cite{Kremer2016}, and in $^{94}$Zr in a neutron scattering experiment~\cite{Chakraborty2013}. Similarly, coexistence of deformed and spherical configurations was also observed in $^{96,98}$Sr isotopes in a Coulomb excitation experiment performed at ISOLDE~\cite{Clement2016_PRL,Clement2016_PRC}.

Already in the late 1970s, it has been pointed out by Federman and Pittel that the strong isoscalar attractive proton-neutron interaction between the spin-orbital partners, in particular, the $\pi(1g_{9/2})$ and the $\nu(1g_{7/2})$, could be responsible for the rapid emergence of deformation~\cite{Federman1977,Federman1978,Federman1979}. As neutrons are added beyond the $\nu(d_{5/2})$ orbital, the $\nu(1g_{7/2})$ will also start to fill. Due to the strong isoscalar interaction, the gap between the $\pi(1g_{9/2})$ and the $\pi(2p_{1/2})$ is reduced, which makes it energetically favorable for protons to be promoted from the $\pi(2p_{1/2})$ into the  $\pi(1g_{9/2})$ orbital. The filling of the  $\pi(1g_{9/2})$ orbital, successively, lowers the  $\nu(1g_{7/2})$, further promoting neutrons into it. The breaking down of the pairing  $\pi$-$\pi$ and $\nu$-$\nu$ correlations and the development of spatial $\pi$-$\nu$ correlation leads to deformation. This qualitative description was also supported by HFB and shell-model calculations, though in a very limited valence space from the current perspective, using a $^{94}$Sr core~\cite{Federman1979}. The calculations also showed that the first excited $0^+$ state in $^{98}$Zr is strongly mixed and is thus possibly deformed. It is important to point out that the specific ordering of the orbitals around A=100 makes this effect very strong allowing for the rapid onset of collectivity.

The microscopic origin of the strong interaction between the spin-orbit partner orbitals can be understood in terms of the tensor-force component of the nucleon-nucleon interaction~\cite{Otsuka2005,Otsuka2010}, which is a direct consequence of its meson exchange character. The importance of the tensor force in the shell-evolution has been outlined in Ref.~\cite{Otsuka2016}. In the same publication, the authors have also stressed the importance of particle-hole excitations in the evolution of the shell structure especially their role in the transition probabilities.
Indeed, the recently performed large-scale shell-model calculations, which do not take particle-hole excitations into account, carried out for Zr isotopes of  $N$~=~50-60 with a $^{78}$Ni core, were able to account for the sudden drop in the excitation energy of the first excited $2^+$ states at $N$ = 60, but were unable to correctly describe the abrupt rise of the transition probabilities~\cite{Sieja2009}. The recent advances of the Monte-Carlo shell-model calculations~\cite{Noritika2012}, have allowed Togashi et al.~\cite{Togashi2016} to perform calculations for the Zirconium isotopes of $N$~=~50-70 with a much larger basis, including also neutron excitation across the $N$~=~50 shell closure. The calculation reproduces both the rise of the $B(E2)$ values and the drop in the energies of the first excited states along the isotopic chain. 
The MCSM calculations also predict a shape coexistence of more than two configurations  with different deformations in the region around $N=60$. Similarly, HFB calculations for $^{98}$Zr based on the VAMPIR model~\cite{Petrovici2012}, predict a coexistence of several strongly mixed shapes, albeit, with noticeable discrepancies with respect to the data on some electromagnetic properties.
In another approach, the shape-transition in the Zr isotopes was discussed in the framework of configuration mixing in the interacting boson model (IBM-CM)~\cite{Gavrielov2019,Gavrielov2020,Garcia-Ramos2019}. The calculation in Refs.~\cite{Gavrielov2019,Gavrielov2020} suggests the so-called intertwined quantum phase transitions, which involves crossing of two configurations, where each of the two configurations undergoes its own quantum phase transition.

The $^{98}$Zr nucleus lies on the interface between the spherical and the deformed region making it pivotal to understanding shape transition and the shape coexistence in the A $\approx$ 100 region. Very recently, the lifetimes of the yrast $2^+$ and $4^+$ were determined by the recoil-distance Doppler-shift (RDDS) technique in a fission experiment at GANIL~\cite{Singh2018}, but the lifetimes of the second excited $2^+$ and $4^+$ states remain unknown up to today. In this article we report on a measurement of the lifetimes of the $2^+_1$, $2^+_2$, $2^+_3$, $4^+_1$, $4^+_2$ states. Additionally, the lifetime of the $3^-_1$ state has been measured. 
\section{Experiment}
The nucleus of interest was populated in the $^{96}$Zr($^{18}$O,$^{16}$O)$^{98}$Zr two-neutron transfer reaction.
An average beam current of 1~pnA with an energy of 50~MeV
was provided by the Cologne 10 MV FN-Tandem accelerator. A 1~mg/cm$^2$ $^{96}$Zr self-supporting foil enriched to 72.47~\% was stretched inside the Cologne Plunger device~\cite{Dewald2012}. To stop the nuclei ejected after the transfer reactions induced on the target, a 6.5~mg/cm$^2$ Ta stopper was stretched parallel to the target. The $\gamma$~rays produced in the experiment were detected by 11 high purity germanium (HPGe) detectors positioned in two rings around the target chamber. Five detectors were placed at backward angles of 142$\degree$ relative to the beam axis and six at forward angles of 45$\degree$. Recoiling light fragments were detected by an array of six solar cells placed at backward angles inside the target chamber, covering angles between 120$^o$ and 165$\degree$. The data were recorded at 7 target-to-stopper distances (22 $\mu$m, 41 $\mu$m, 71 $\mu$m, 101  $\mu$m, 131 $\mu$m, 221 $\mu$m, 321 $\mu$m) in triggerless mode. These distances were determined relative to a zero point which is obtained by using the capacitive method as described in Ref.~\cite{Alexander1970,Dewald2012}. For each distance and each detector ring particle-$\gamma$ coincidences were sorted off-line.

\begin{figure}[!]
\includegraphics[width=\linewidth]{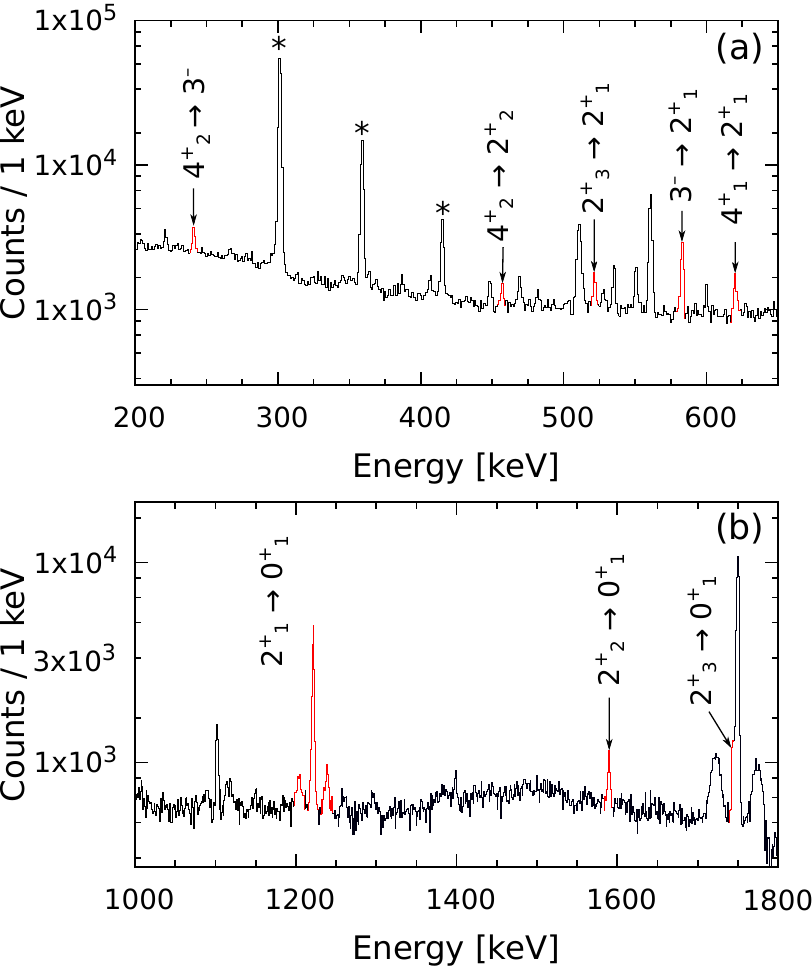}
\caption{Particle-gated $\gamma$-ray singles spectrum of both detector rings for plunger distance of 22 $\mu$m. (a) and (b) show different cutouts of the spectrum. The transitions belonging to $^{98}$Zr are indicated and colored in red. The peaks belonging to Coulomb excitation in the $^{181}$Ta stopper are indicated with an asterisk.}
\label{fig:spectrum}
\end{figure}

The particle-gated spectrum for the smallest distance of 22 $\mu$m is shown in Fig.~\ref{fig:spectrum}. Due to the low angular granularity of the solar cells and the straggling of the recoiling nuclei out of the target, no clear separation between $^{16}$O and $^{18}$O could be achieved in the particle spectrum. Hence, the major peaks in the $\gamma$-ray spectrum are due to Coulomb excitation in the target and the stopper foils. Transitions belonging to $^{97}$Zr and $^{100}$Mo are also observed, populated in the single-neutron and the alpha-transfer reactions, respectively. Despite the presence of many transitions, the ones belonging to $^{98}$Zr are well defined and are indicated in Fig.~\ref{fig:spectrum}. The transition intensities have been measured by integration and were normalized to the intensity of the 2$_1^+ \rightarrow$ 0$_1^+$ transition. Additionally, weak transitions from the $0^+_3$ and the $0^+_4$ states are observed. The intensities of the 0$_3^+ \rightarrow$ 2$_1^+$ and the 0$_4^+ \rightarrow$ 2$_2^+$ are very low and comparable with the level of the background fluctuation, i.e. 1 \% of the 2$_1^+ \rightarrow$ 0$_1^+$  transition intensity. The experimental information on the observed $\gamma$-ray transitions is summarized in Table~\ref{t:intensities}.

\begin{table}
\setlength{\tabcolsep}{1pt}
\caption{Relative transition intensities observed in the experiment normalized to the  $2_1^+ \rightarrow 0_1^+$ transition. The energies are taken from Ref.~\cite{NuDat_A98}.}
\label{t:intensities}
\begin{ruledtabular}
\begin{tabular}{ccc}
Transition & Transition energy [keV] & Intensity   \\ \hline
 $2_1^+ \rightarrow 0_1^+$ & 1223 & 100.0(37) \\
 $0_3^+ \rightarrow 2_1^+$ & 213 &  1.0(10)\\
 $2_2^+ \rightarrow 0_1^+$ & 1591 & 10.4(14) \\
 $2_3^+ \rightarrow 0_1^+$ & 1744 &  23.5(50)  \\
 $2_3^+ \rightarrow 2_1^+$ & 522 & 5.2(10) \\
 $3^-_1 \rightarrow 2_1^+$ & 583 & 17.7(12) \\
 $4_1^+ \rightarrow 2_1^+$ & 621 & 9.0(12) \\
 $0_4^+ \rightarrow 2_2^+$ & 269 & 1.0(10)\\
 $4_2^+ \rightarrow 3^-_1$ & 242 & 10.1(12) \\
 $4_2^+ \rightarrow 2_2^+$ & 456 & 3.1(9) \\ 
 $4_2^+ \rightarrow 4_1^+$ & 204 & 3.1(16) \\
 $4_2^+ \rightarrow 2_1^+$ & 825 & 2.9(9) \\  
\end{tabular}
\end{ruledtabular}
\end{table}

\begin{figure}[h!]
\includegraphics[width=\linewidth]{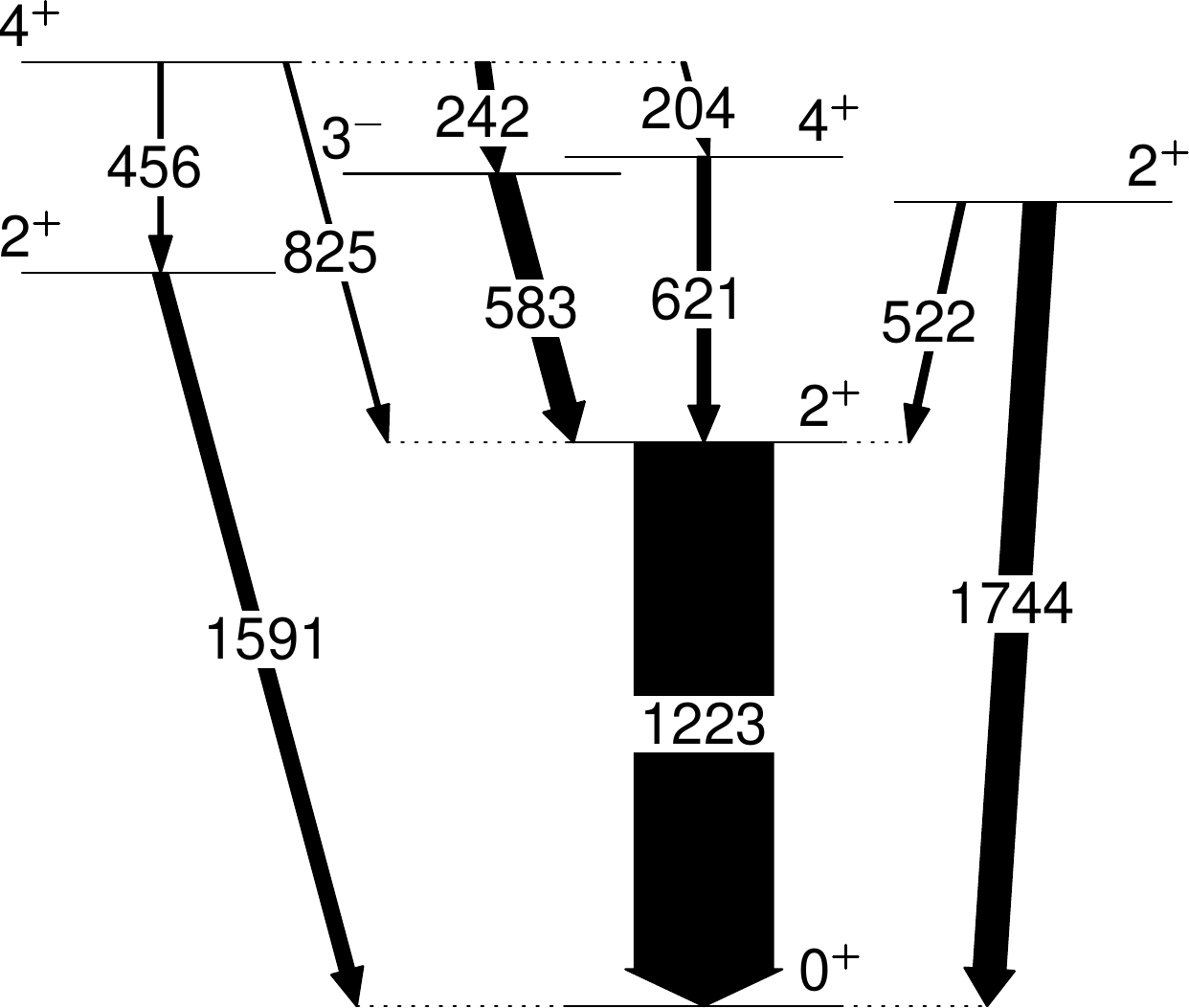}
\caption{Level scheme of $^{98}$Zr populated in the $^{96}$Zr($^{18}$O,$^{16}$O)$^{98}$Zr two-neutron transfer reaction at beam energy of 50 MeV. The width of the lines are proportional to the transition intensities given in Table~\ref{t:intensities}. }
\label{fig:scheme}
\end{figure}

\begin{figure}[h!]
\includegraphics[width=\linewidth]{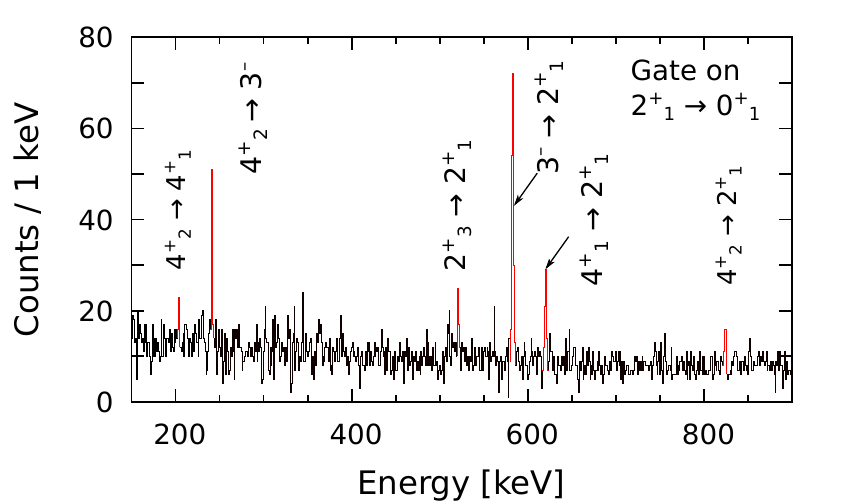}
\caption{Particle-$\gamma$ gated $\gamma$-ray spectrum of both detector rings for plunger distance of 22 $\mu$m. The transitions belonging to $^{98}$Zr are indicated and colored in red.}
\label{fig:gated}
\end{figure}

Using $\gamma$-$\gamma$ coincidences, a level scheme has been built and is shown in Fig.~\ref{fig:scheme}. The spectrum gated on the $2_1^+ \rightarrow 0_1^+$ transition is displayed in Fig.~\ref{fig:gated}. This spectrum is also used to check for other feeding contributions not clearly observed in the singles spectrum. Additionally, this spectrum allows for a cross-check of the intensities obtained using the singles $\gamma$-ray spectrum, by comparing the ratio of the intensities obtained in the singles and the gated spectrum. These ratios are consistent within the experimental uncertainties. 

\section{Lifetime determination and results}
To extract the lifetimes of the $4^+_2$,$4^+_1$,$3^-_1$,$2^+_2$ and $2^+_1$ states, the RDDS technique has been used. The data has been analyzed using the Bateman equations (BEs) and the differential decay curve method (DDCM)~\cite{Dewald1989}. Here we present only the essential ideas needed for the analysis. For a detailed review of both methods the reader is referred to Ref.~\cite{Dewald2012}. For the sake of clarity, we use the same notation as in Ref.~\cite{Dewald2012}.

An excited state $i$ of a nucleus ejected from the target foil can decay either in-flight or after stopping in the stopper foil. The $\gamma$ rays emitted by a nucleus in-flight would appear Doppler shifted in the $\gamma$-ray spectrum.
The corresponding peak is known as the \textit{shifted} peak and its intensity, \textit{i.e.} number of counts, is given by $I^s_{i}(t)$, where $t$ is the time of flight of the nucleus between the target and the stopper. If the $\gamma$ decay occurs after the nucleus has stopped in the stopper foil the peak in the $\gamma$-ray spectrum would not experience a Doppler shift and is known as the \textit{unshifted} peak and its intensity is given by $I^u_{i}(t)$. The so-called decay curve is defined as:

\begin{equation}
R_i(t)=I_i^u(t)/ \left( I_i^s(t)+I_i^u(t) \right) .
\end{equation}
In the case where state $i$ is not fed from another state,\textit{i.e.} is directly populated in a nuclear reaction, the decay curve is given by the simple formula:
\begin{equation}
\label{eq:simple}
R_i(t)=e^{-t\lambda_i},
\end{equation} 
where $\lambda_i$ is the decay constant of state $i$ and is related to the level lifetime $\tau_i$ with $\lambda_i=1/\tau_i$.
In a realistic case, the excited state has a complicated feeding pattern. The feeding contributions need to be taken into account to obtain the correct lifetime. One needs to solve the Bateman equations, which are a system of first-order differential equations that relate the populations $n_i(t)$ of the excited states $i$ as a function of the time $t$, depending on the decay constants $\lambda_i$ of the states $i$ and the branching ratios. The Bateman equations are:
\begin{equation}
\frac{d}{dt}n_i(t)=-\lambda_i n_i(t)+\sum_{k=i+1}^{K}\lambda_k n_k(t)b_{ki}.
\end{equation}   
Here $k$ denotes the excited states feeding the state $i$, $b_{ki}$ are the branching ratios between states $k$ and $i$ and $K$ is the total number of states.
The solutions of these equations with respect to the decay curves $R_i(t)$ is given by:
\begin{equation}
\label{eq:Bateman}
R_i(t)=P_ie^{-t\lambda_i}+\sum_{k=i+1}^{K}M_{ki} \left[(\lambda_i/\lambda_k)e^{-t\lambda_k}-e^{-t\lambda_i} \right].
\end{equation}
$M_{ki}$ is defined recursively as:
\begin{equation}
\begin{split}
&M_{ki}(\lambda_i/\lambda_k -1)= \\
&b_{ki}P_k-b_{ki}\sum_{m=k+1}^{K}M_{mk}
+\sum_{m=i+1}^{k-1}M_{km}b_{mi}(\lambda_m/\lambda_k),
\end{split}
\end{equation} 
where $P_i$ is the population of the state $i$.
Finding a solution to these equations becomes a tedious task, prone to errors when the feeding pattern is complicated like in the case of compound and fission reactions, where the spin and the energy transfer of the reaction are high. However, due to the low-spin and low-energy transfer in the 2n-transfer reaction, relatively few states are populated. Moreover, these states are populated directly, not through a compound state. In such a case, the feeding pattern is simple, making the direct application of Eq.~\ref{eq:Bateman} to the experimental data relatively straightforward.

\begin{figure*}[h!]
\includegraphics[width=\linewidth]{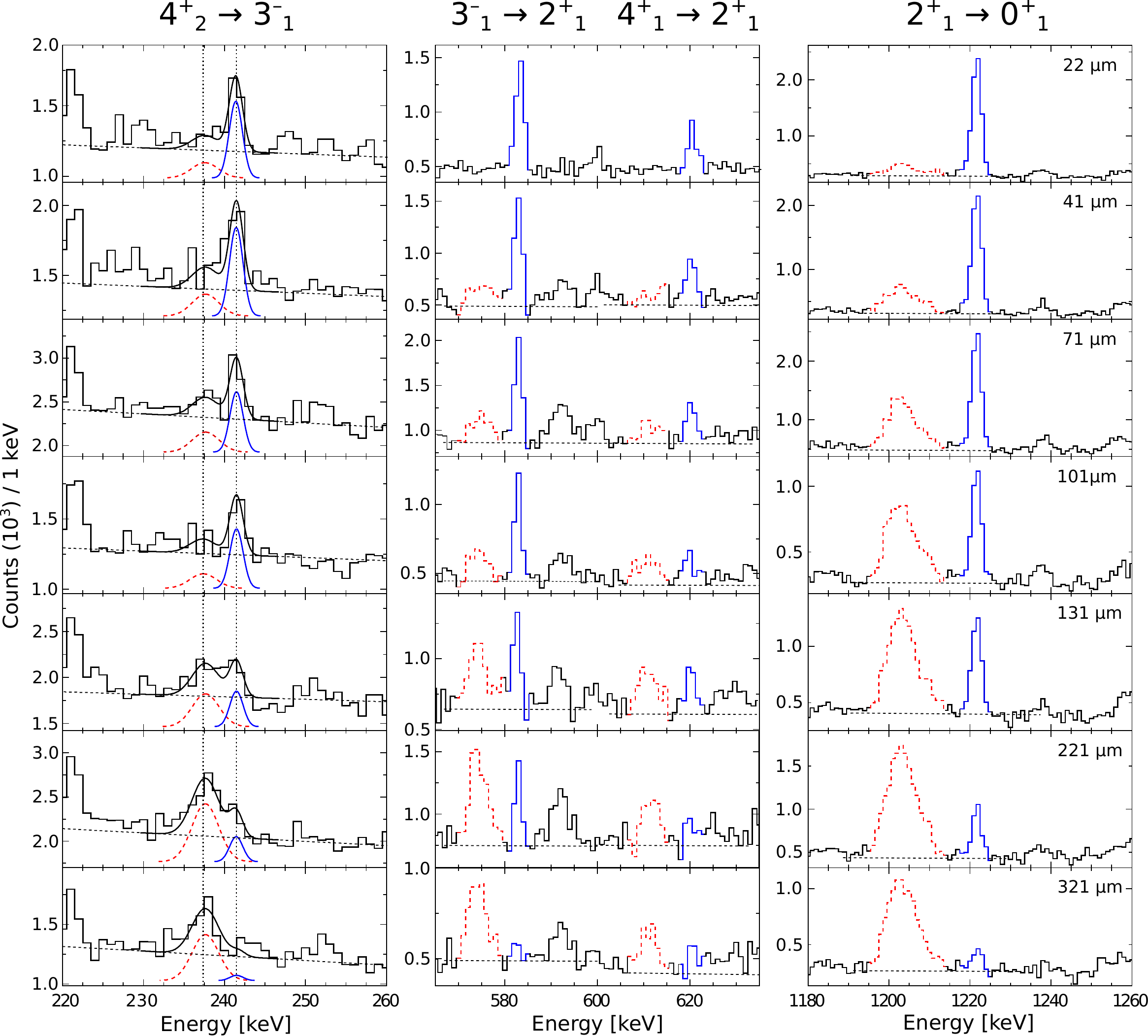}
\caption{Particle-gated spectra of the backward detector ring for all distances (indicated in the top-right corner) of the $4^+_2 \rightarrow 3^-_1$ (left), the $3^- \rightarrow 2^+_1$ and the $4^+_1 \rightarrow 2^+_1$ (middle) and the $2^+_1 \rightarrow 0^+_1$ (right) transitions used to obtain the shifted (dashed-red) and the unshifted (blue) intensities. The horizontal dashed lines represent the level of the background used for the determination of the peak areas were obtained.}
\label{fig:Plunger}
\end{figure*}

To conduct the RDDS analysis for each distance, two particle-gated spectra were generated, one for each ring. Drifts induced by radiation damage in the solar cells were compensated by a shift-tracking procedure.

The particle-gated spectra of the backward ring for each distance are displayed in Fig.~\ref{fig:Plunger}. One can clearly see the evolution of the shifted components for the states of interest. In all the cases except the $4^+_2 \rightarrow 3^-_1$ transition, the shifted and the unshifted peaks are well separated and their areas were determined by integration. The advantage of integration is that, no assumptions on the shape of the peaks are made and hence possible systematic errors are reduced. The systematic error that  arises when choosing the fit region and the background parametrisation has been take into account when obtaining the uncertainties of the $R_i(t)$ values. The $4^+_2 \rightarrow 3^-_1$ transition has an energy of 242 keV and the shifted and unshifted components are not well separated. To obtain the areas of the two components, a fit to the spectra have performed a using two Gaussians, keeping the peak positions and the widths of both components fixed for all the distances.

The average speed of the ejected $^{98}$Zr nuclei was determined directly from the spectra, by measuring the energy difference between the shifted and the unshifted peaks for the strongest observed transitions, \textit{i.e.} $2_1^+ \rightarrow 0_1^+$ and the $3^-_1 \rightarrow 2_1^+$ transitions, using both the forward and backward angles. All four results were consistent. The average velocity was adopted as 1.89(6)\%~c. Using this velocity we have determined the average time of flight of the $^{98}$Zr nuclei between the target and the stopper for each  distance and have used these values in the following analysis.

\subsection{Bateman equations analysis}
When performing the analysis using the Bateman equations (Eq.~\ref{eq:Bateman}) for a certain level, the level lifetime is used as the only fit parameter. The $\gamma$-ray transition intensities used in equations are the ones from Table~\ref{t:intensities}. A top-to-bottom approach was adopted where the lifetimes of the highest states are determined first and are used as fixed parameters when determining the lifetimes of the lower lying states.

\begin{figure}[h!]
\includegraphics[width=\linewidth]{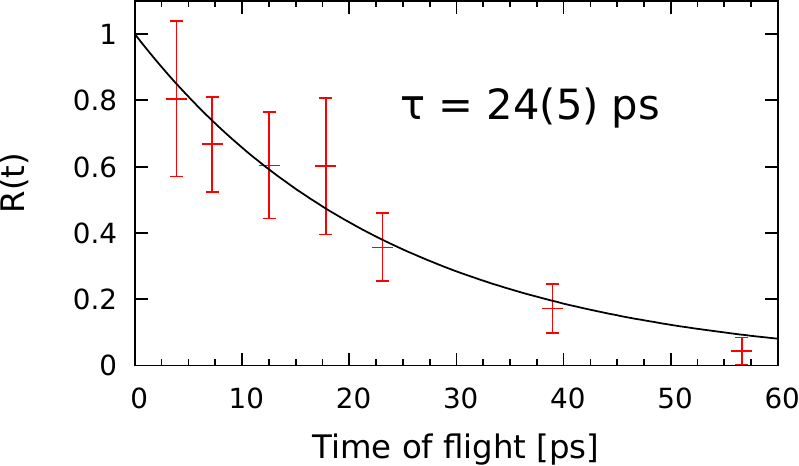}
\caption{Fitted decay curve together with the data points for the $4^+_2$ state using the $4^+_2 \rightarrow 3^-_1$ transition observed at backwards angle. The obtained lifetime is given as well.}
\label{fig:242}
\end{figure}

The $4_2^+$ state has no observed feeders and one can simply use Eq.~\ref{eq:simple} to determine the lifetime of the state. A simple exponential decay fit yields a lifetime of $\tau_{4_2^+}$ = 24(5) ps. The fit and the data points are displayed in Fig.~\ref{fig:242}. Due to the much larger background present in the forward detector ring, especially at the low energies, an analysis for this ring was not possible.

\begin{figure}[h!]
\includegraphics[width=\linewidth]{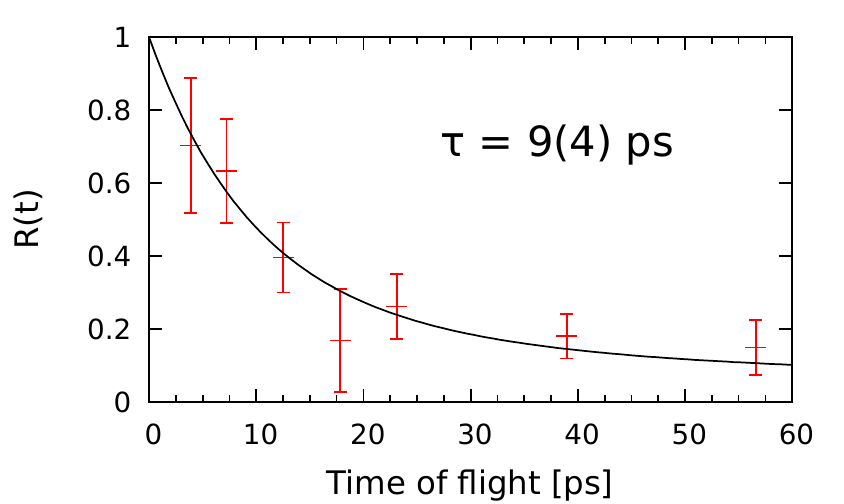}
\caption{Same as Fig.~\ref{fig:242} but for the $2^+_2 \rightarrow 0^+_1$ transition.}
\label{fig:1591}
\end{figure}

The lifetime $\tau_{4_2^+}$ is then used as a fixed parameter when determining the lifetime of the $2^+_2$ state. Here, also the long feeding coming from the $0^+_4$ state is taken into account.
Using the lifetime $\tau_{2_2^+}$ as the only fit parameter in Eq.~\ref{eq:Bateman} the data points for the decay curve of the $2_2^+$ state were fitted, resulting in a lifetime of $\tau_{2^+_2}$ = 9(4) ps. The fit to the data points is displayed in Fig.~\ref{fig:1591}.

\begin{figure*}[h!]
\includegraphics[width=\linewidth]{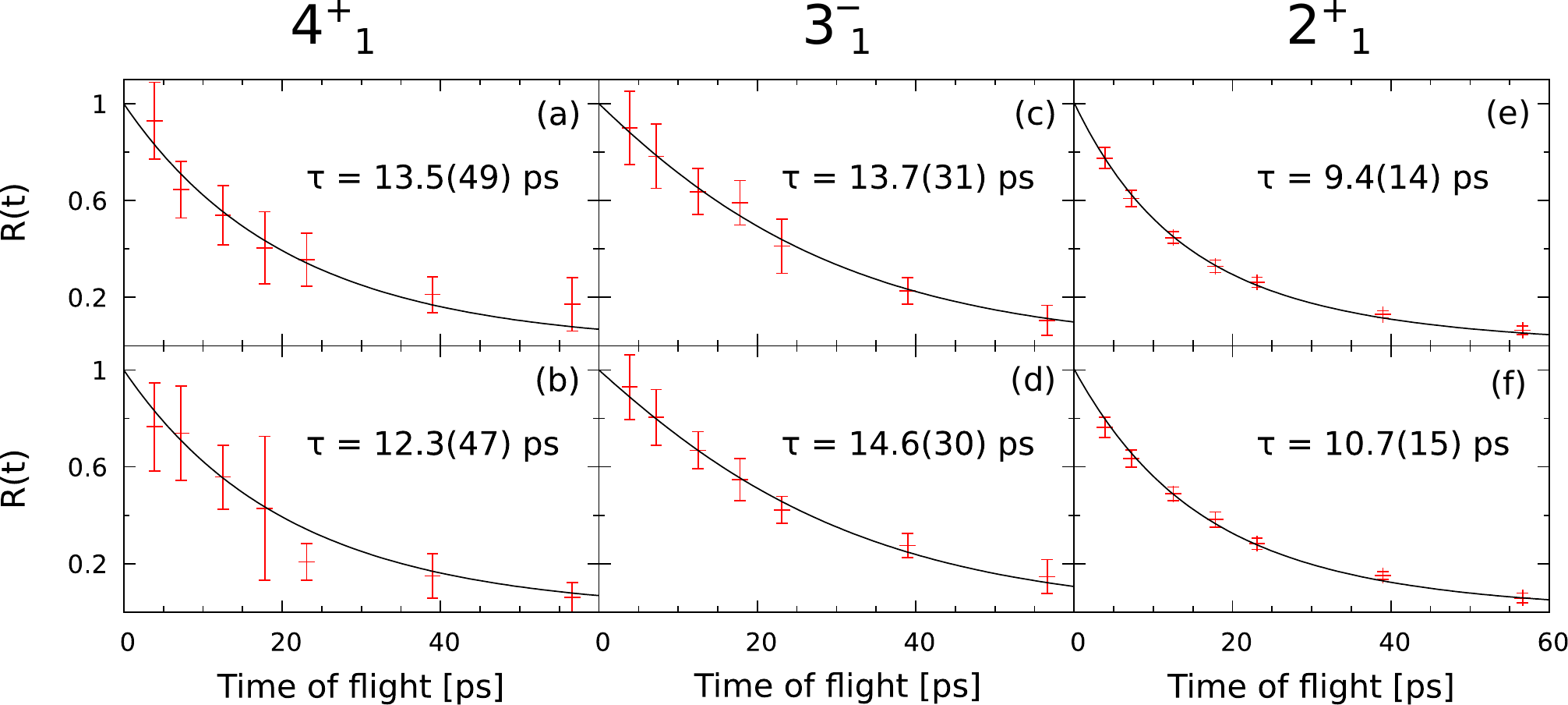}
\caption{Fitted decay curve together with the data points: (a) and (b) for the $4^+_1$ state, (c) and (d) for the $3^-_1$ and (e) and (f) for $2^+_1$. The upper panel is for the backward detector ring and lower for the forward detector ring.}
\label{fig:Bateman}
\end{figure*}

The lifetime of the $4^+_1$, the $3^-_1$ and the $2^+_1$ states were obtained using the same procedure.  The fits to the data and the obtained lifetimes are shown in Fig.~\ref{fig:Bateman}.

The experimental uncertainties of the measured lifetimes were obtained by performing a Monte-Carlo simulation, similarly to~Ref.~\cite{Litzinger2015}. All the input parameters used in the fit, are independently varied within the corresponding experimental uncertainties, before performing the fit. Since the individual fit values are distributed symmetrically around a mean value, the error is defined simply as the standard deviation of these values.

\subsection{DDCM analysis}
\label{sec:DDCM}
The data for the $2^+_1$, $3^-$ and $4^+_1$ states were also analyzed by the DDCM. This method was developed by Dewald et al. in 1989~\cite{Dewald1989} and is derived from the Bateman equations. The method is transparent and easier to apply when the feeding pattern is complicated. One of the advantages of DDCM is that it relies on the relative distances between the target and the stopper, which are very precisely determined via the active feedback system of the Plunger device~\cite{Dewald2012}.  
In the framework of the DDCM the lifetime can be derived for each distance from:
\begin{equation}
\tau_i(x)=\frac{R_i(x)-\sum_{k}^{}(b_{ki}I_k(x)/I_i(x))R_{k}(x)}{v\frac{d}{dx}S_i(x)}=\frac{U_i(x)}{v\frac{d}{dx}S_i(x)},
\end{equation} 
where $I_i$ and $I_k$ are the total intensities of the transitions depopulating the states $i$ and $k$ and $v$ is the speed of the ejected nuclei. The sum is carried over all the feeders $k$ of the state of interest $i$. The numerator can be interpreted as a decay curve which has been corrected for the "stopped" feeding. The term in the denominator is the shifted component normalized to the total intensity:
\begin{equation}
S_i(x)=\frac{I_i^s(x)}{I_i^s(x)+I_i^u(x)}.
\end{equation}

\begin{figure*}[h!]
\includegraphics[width=\linewidth]{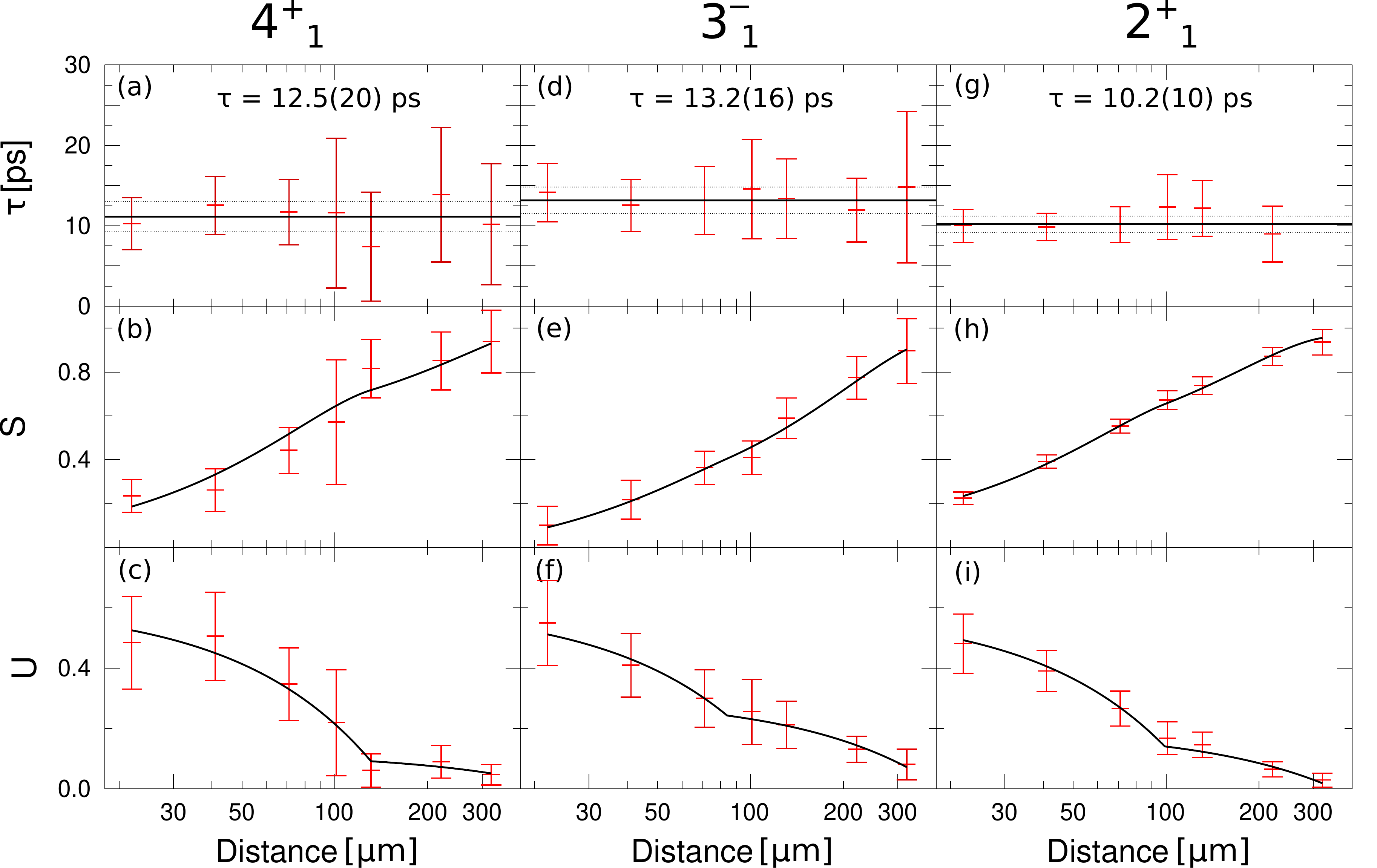}
\caption{DDCM analysis (see text) of the backward-detector ring for the $4^+_1$ state (a)-(c), for the $3^-_1$ state (d)-(f) and for the $2^+_1$ state (g)-(i). The individual lifetimes obtained in the analysis are displayed in the upper panel. The middle panel shows the evolution of the shifted component and the fit to it used to obtain the derivative $d(S_i(x))/dx$. The lower panel shows the evolution of the unshfited component and the fit to it used to obtain the individual lifetimes.}
\label{fig:ddcm}
\end{figure*}

The data were analyzed using the computer code NAPATAU which is described in Ref.~\cite{Saha2004}.
The program performs a piecewise polynomial fit to the shifted intensities $S_i(x)$ to obtain the derivative $d(S_i(x))/dx$. The derivative is multiplied by a parameter $\tau_i(x)$ and the product $\tau_i(x)d (S_i(x))/dx$ is fitted simultaneously to the feeder-corrected decay curve values $U_i(x)$. The fit, in this case, has been performed with two second-order polynomials. The parameters $\tau_i(x)$ are the lifetime of the state $i$ by definition. The final value of $\tau_i$ is obtained as a weighted average of the individual results. Fits to the data of the backward-detector ring used to extract the lifetimes of the $4^+_1$, the $3^-$ and the $2^+_1$ states are shown in Fig.~\ref{fig:ddcm}. The same procedure is performed for the forward-detector ring as well. The results of the backward detector ring are given in Table.~\ref{t:results}.

\subsection{DSA analysis of the $2^+_3$ state}
\label{DSA}
\begin{figure}[h!]
\includegraphics[width=\linewidth]{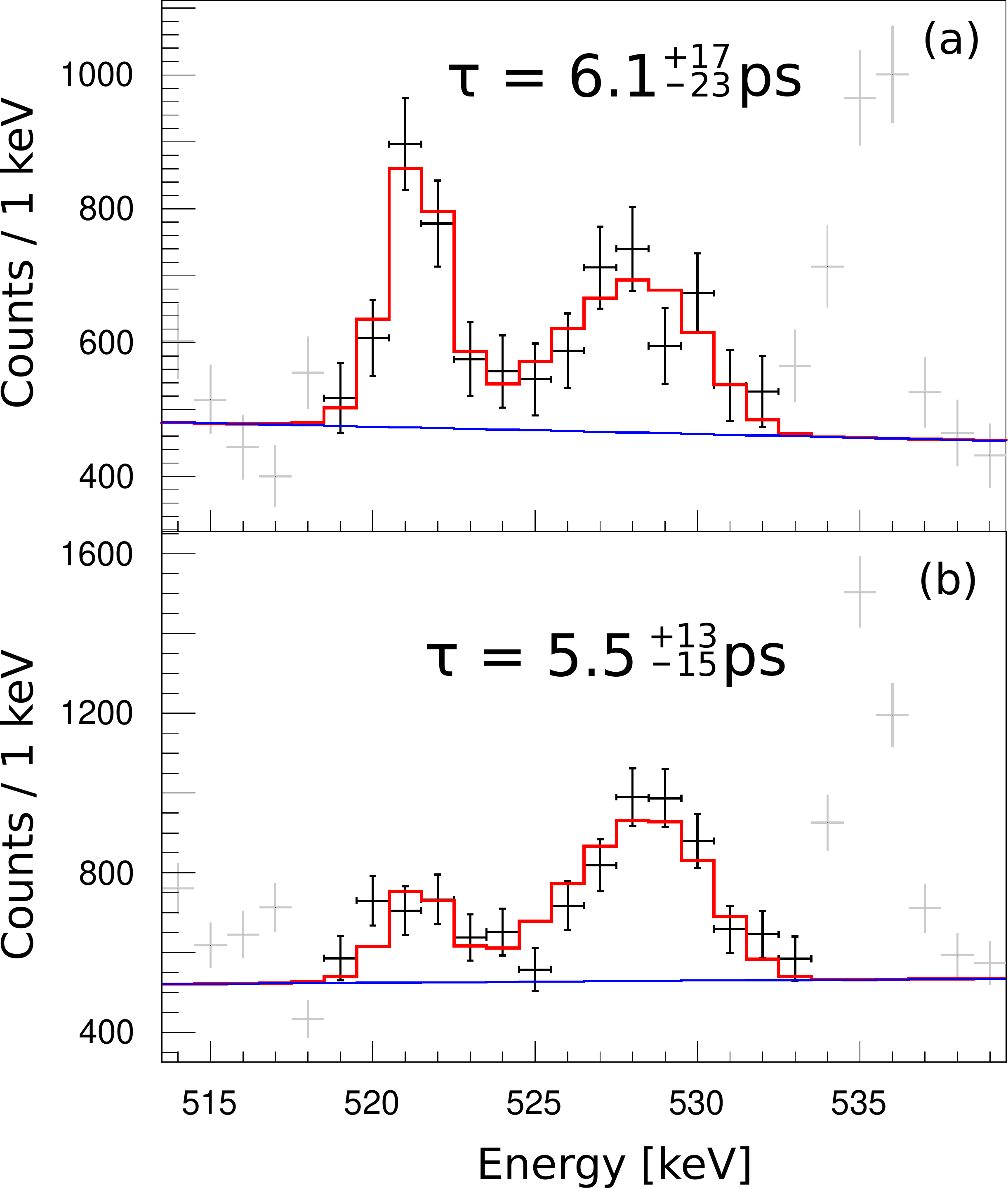}
\caption{DSA fits to the Doppler-broadened lineshapes of the $2^+_3 \rightarrow 2^+_1$ transition for the (a) 22 $\mu$m and the (b) 41 $\mu$m distances of the forward detector ring  used to obtain the lifetime of the $2^+_3$ state.}
\label{fig:2p3}
\end{figure}

When the lifetimes of the states are comparable to the average stopping time of the ejecting nuclei inside the stopper, the RDDS technique cannot be applied directly. To obtain the correct lifetime, the decays that happen during the stopping needs to be taken into account by employing the Doppler-shift attenuation (DSA) method. For a detailed review of the method, the reader is referred to Ref.~\cite{Fossan1974,Nolan1979}. To determine the lifetime of the $2^+_3$ state, a DSA analysis has been performed utilizing the program APCAD~\cite{Stahl2017}. In APCAD, the slowing down of the $^{98}$Zr ions inside the target and the stopper and the drift between them is modeled using a Monte-Carlo simulation in the framework of Geant4~\cite{Geant4}. The electronic and the nuclear stopping powers used in the simulation are taken from SRIM2013~\cite{SRIM}. The doubly-differential cross-section of the  $^{96}$Zr($^{18}$O,$^{16}$O)$^{98}$Zr reaction used in the simulation has been calculated using the GRAZING code~\cite{Winther1993,Yanez2015}. 
After the traces of the individual ions are simulated, APCAD calculates the Doppler shift observed in the individual detectors as a function of the time after the production of the $^{98}$Zr nuclei inside the target. The calculations take into account the setup geometry, the kinematic restrictions imposed by the solar cells and the detector resolutions. 
Finally, the simulated Doppler–broadened $\gamma$-ray lineshapes are fitted to the experimental spectra using the level lifetime as the only fit parameter.
The fit has been performed to the $2^+_3 \rightarrow 2^+_1$ transition peak. Only the forward detector ring spectra were used since the shifted component of the peak in the backward detector ring spectrum lies in the tail of the 511 keV peak.
The fit to the spectra for the 22 $\mu$m and the 41 $\mu$m distances are shown in Fig.~\ref{fig:2p3}.
The errors indicated in the figure include the statistical error of the fit and the systematic errors assuming 10$\%$ uncertainty in the stopping powers and 5 $\mu$m uncertainty in the distance between the target and the stopper. Additionally, up to 10 $\%$ long-lived feeding has been assumed as a possible source of systematic error.

The results from the lifetime measurements are summarized in Table~\ref{t:results}. Adopted values are given and are compared to the recent results from Ref.~\cite{Singh2018}. While the value for the $4_1^+$ that we report here is consistent with the one from Ref.~\cite{Singh2018}, the lifetime of the $2^+_1$ we measure is considerably longer. It should be noted that the lifetimes of Ref.~\cite{Singh2018} have been determined also in a singles RDDS analysis, but the nucleus was populated in a fission reaction, which makes the feeding pattern severely complicated. 
Using the measured lifetimes, and the information on the conversion coefficients, the multipolarity mixing ratios and the branching ratios from Ref.~\cite{NuDat_A98}, the reduced transition probabilities are calculated and the results are presented in Table~\ref{t:be2}. One can attempt to use the branching of the $2^+_2 \rightarrow 0_3^+$ transition, measured in Ref.~\cite{Becker1984} and evaluated in Ref.~\cite{NuDat_A98}, in order to estimate the reduced transition probability. However, this leads to an unrealistically large $B(E2)$ value and, furthermore, the indicated branching ratio was not confirmed by the more recent measurement in Ref.~\cite{Urban2017}. Accordingly, the lifetime of the $2^+_2$ and the branching of the $2^+_2 \to 0^+_3$ needs to be further corroborated in order to pin down an accurate $B(E2)$ value.

\begin{table}
\setlength{\tabcolsep}{1pt}
\caption{Lifetimes measured in the experiment using the BEs, the DDCM and the DSA method together with the adopted values. The results from Ref.~\cite{Singh2018} are given for comparison.}
\label{t:results}
\begin{ruledtabular}
\begin{tabular}{c|cc|cc|c|c}

\multicolumn{7}{c}{Lifetime [ps]}\\
\hline
 &\multicolumn{2}{c|}{Backward ring}&\multicolumn{2}{c|}{Forward ring}& & \\
 \hline
 State&BE&DDCM&BE&DDCM& \textbf{Adopt.} & Lit.\tablenotemark[1]\\

$2_1^+$ & 9.4(14) &10.2(10)&  10.7(15) & 10.4(10) &\textbf{10(2)} &3.8(8) \\
$2_2^+$ & 9(4) &---& --- & --- &\textbf{9(4)} &--- \\
$3_1^-$ & 13.7(31)& 13.2(16)&14.6(30) & 13.2(16) & \textbf{14(3)}&---  \\
$4_1^+$ &13.5(49)&12.5(20) &12.3(47)& 11.5(19) & \textbf{13(5)} & 7.5(15)\\
$4_2^+$ &24(5)&---&---&---&\textbf{24(5)}&--- \\
\hline
&&&\multicolumn{2}{c|}{DSA} & \\
\hline
&&&22 $\mu$m&41 $\mu$m&&\\

$2_3^+$&&& $6.1^{+1.7}_{-2.3}$&$5.5^{+1.3}_{-1.5}$&$\mathbf{6^{+2}_{-3}}$&---\\

\end{tabular}
\end{ruledtabular}
\begin{flushleft}
	\tablenotemark[1] From Ref.~\cite{Singh2018}.
\end{flushleft}
\end{table}

\section{Discussion}
\subsection{Comparison to MCSM calculations}

First, we compare the new measured results with the recent large-scale Monte-Carlo shell-model calculation of Togashi et al. \cite{Togashi2016}. The calculation reproduced  both the drastic rise of the $B(E2)$ values and the drop in the energies of the first excited state in the $N$ = 50-70 zirconium isotopes. The results for $^{98}$Zr are presented in Table~\ref{t:be2}. We point out different assignments to experimental states of the calculated levels in $^{98}$Zr. In the first comparison done in Ref.~\cite{Ansari2017} (MCSM-1), the first excited 4$^+$ and 6$^+$ were assigned to the calculated first excited 4$^+$ at 1.59 MeV and 6$^+$ at 1.64 MeV. However, in the more recent work by ~\cite{Singh2018}, these assignments were changed and instead the experimental 4$^+$ and 6$^+$ were assigned to the calculated second excited 4$^+$ at 2.197 MeV and 6$^+$ at 2.668 MeV (MCSM-2). This change improves the agreement for the energies of the states, however it yields a very low $B(E2;4_1^+ \rightarrow 2_1^+$) value of 0.6 W.u. This change does not have such a large effect for the  $B(E2;6_1^+ \rightarrow 4_1^+)$ transition strength as it changes the calculated value from 102 W.u to 87 W.u., while the experimental value has been measured as 103.0(35.7) W.u. in Ref.~\cite{Singh2018}. 

Taking into account the new experimental $B$($E2$) values of the first and second 4$^+$ states, the situation is still unclear. While the $B(E2;4^+_1 \rightarrow 2_1^+)$ value is best reproduced using the old assignment, the $B(E2;4^+_1 \rightarrow 2_2^+)$ value agrees better with the new assignment. As mentioned in Ref.~\cite{Singh2018}, the reason might be related to an underestimation of the 
mixing of the $2^+$ states indicating the need for further refinements 
of the shell-model Hamiltonian. Note that both 2$^+$ states have a small 
ground-state decay rate in this theory and this experiment.

\begin{table*}
\begin{center}
\caption{Experimentally deduced transition probabilities for $^{98}$Zr from the current experiment and from Ref.~\cite{NuDat_A98,Singh2018}, in comparison with theoretical calculations. The conversion coefficient, the multipolarity mixing ratios, and the branching ratios are taken from Ref.~\cite{NuDat_A98} if not otherwise mentioned.}
\label{t:be2}
\begin{ruledtabular}
\begin{tabular}{lcccccc}

\multicolumn{6}{c}{$B(E2)$ [W.u.]}\\
Transition & This work & Singh \textit{et al.}\footnote{From \cite{Singh2018}.\label{foot:singh}}  & MCSM-1\footnote{Calculation from \cite{Togashi2016}. Level assignments as in \cite{Ansari2017}.} &  MCSM-2\footnote{Calculation from \cite{Togashi2016}. Level assignments as in \cite{Singh2018}.\label{foot:togashi-1}} & IBM-CM-1\footnote{Calculation from \cite{Gavrielov2019,Gavrielov2020}.\label{foot:gav}}&  IBM-CM-2 \footnote{Calculation from~\cite{Garcia-Ramos2019}.} 
\\
\hline
$\be{2}{1}{0}{1}$ & $1.1^{+0.3}_{-0.2}$ & 2.9(6) & 0.0 & 0.0 & 1.35 & 9.6 \\
$\be{2}{1}{0}{2}$ & $11^{+3}_{-2}$ & 28.3(6.0) & 70 & 70 & 43.39 & 32 \\
$\be{2}{2}{0}{1}$ & $0.26^{+0.20}_{-0.08}$ & $-$ & 0.0 & 0.0 & 0.34 & 2.5 \\
$\be{2}{2}{0}{2}$ & $1.8^{+1.4}_{-0.6}$ & $-$ & 2.0 & 2.0 & 0.06 & 47\\
$\be{2}{2}{0}{3}$ & $-$ \footnote{See text.} & $-$ & 49 & 49 & 6.54  & 3.2\\
$\be{2}{2}{2}{1}$ & $46^{+35}_{-14}$\footnote{Assuming a pure E2 transition.} & $-$ & 8.7 & 8.7 & 47.22 & 0.55\\
$\be{2}{3}{0}{1}$ & $0.14^{+0.12}_{-0.04}$ & $-$ & $-$ & $-$ & 2.33 & 0.01 \\ 
$\be{2}{3}{0}{2}$ & $1.7^{+1.5}_{-0.5}$ & $-$ & $-$ & $-$ & 2.28 & 0.56 \\ 
$\be{2}{3}{2}{1}$ & $7.6^{+6.5}_{-2.3}$ & $-$ & $-$ & $-$ & 1.81 & 46 \\ 
$\be{4}{1}{2}{1}$ & $25^{+15}_{-7}$ & 43.3(8.7) & 103 & 0.6 & 68.0 & 59 \\
$\be{4}{1}{2}{2}$ & $38^{+26}_{-13}$ & 67.5(13.5) & 0.7 & 76 & 1.68 & 67 \\
$\be{4}{2}{2}{1}$ & $0.6^{+0.17}_{-0.12}$ & $-$ & 0.6 & 103 & $-$ \footnote{Outside IBM-CM model space. See text.\label{foot:ibm}} & 0.05 \\
$\be{4}{2}{2}{2}$ & $4.6^{+1.7}_{-1.3}$ & $-$ & 76 & 0.7 & $-$ \footref{foot:ibm} & 0.11 \\			
$\be{6}{1}{4}{1}$ & $-$& 103.0(35.7)&102 & 87 & 76.9 & 143  \\
\hline
				   & \multicolumn{1}{c}{ENSDF\footnote{From~\cite{NuDat_A98}}} &  \\
$\be{0}{3}{2}{1}$ & \multicolumn{1}{c}{58(8)} &&&&37 & 53\\
$\be{0}{4}{2}{2}$ & \multicolumn{1}{c}{42(3)} &&&&46 & 42 \\
$\be{0}{4}{2}{1}$ & \multicolumn{1}{c}{0.103(8)} &&&& 0.045 &0.33 \\
\end{tabular}
\end{ruledtabular}
\end{center}
\end{table*}

\subsection{Comparison to IBM-CM calculation}\label{sec:ibm-cm}

The framework of the interacting boson model with configuration mixing (IBM-CM)~\cite{Duval1982,Duval1983} was recently employed in a calculation \cite{Gavrielov2019, Gavrielov2020,Garcia-Ramos2019} of several observables for the chain of zirconium isotopes with neutron numbers 52-70. The calculation considers  a $^{90}_{40}$Zr$_{50}$ core with valence neutrons in the 50-82 major shell and two configurations. The normal $A$-configuration ([$N_b$]-boson space) corresponds to no active protons above the $Z\!=\!40$ sub-shell gap and the intruder $B$-configuration ([$N_b+2$]-boson space) corresponds to a two-proton excitation from below to above this gap, creating $2p-2h$ states.
The resulting eigenstates $\ket{\Psi;L}$ with angular momentum $L$ are linear combinations of the wave functions $\Psi_A$ and $\Psi_B$ in the two spaces $[N_b]$ and $[N_b+2]$,
\begin{align}
\ket{\Psi;L} = a\ket{\Psi_A;[N_b],L} + b\ket{\Psi_B;[N_b+2],L},
\end{align}
with $a^{2}+b^{2}=1$ and $N_b=4$ is the appropriate boson number for $^{98}$Zr.

\begin{figure*}[h!]
\centering
\begin{overpic}[width=0.49\linewidth]{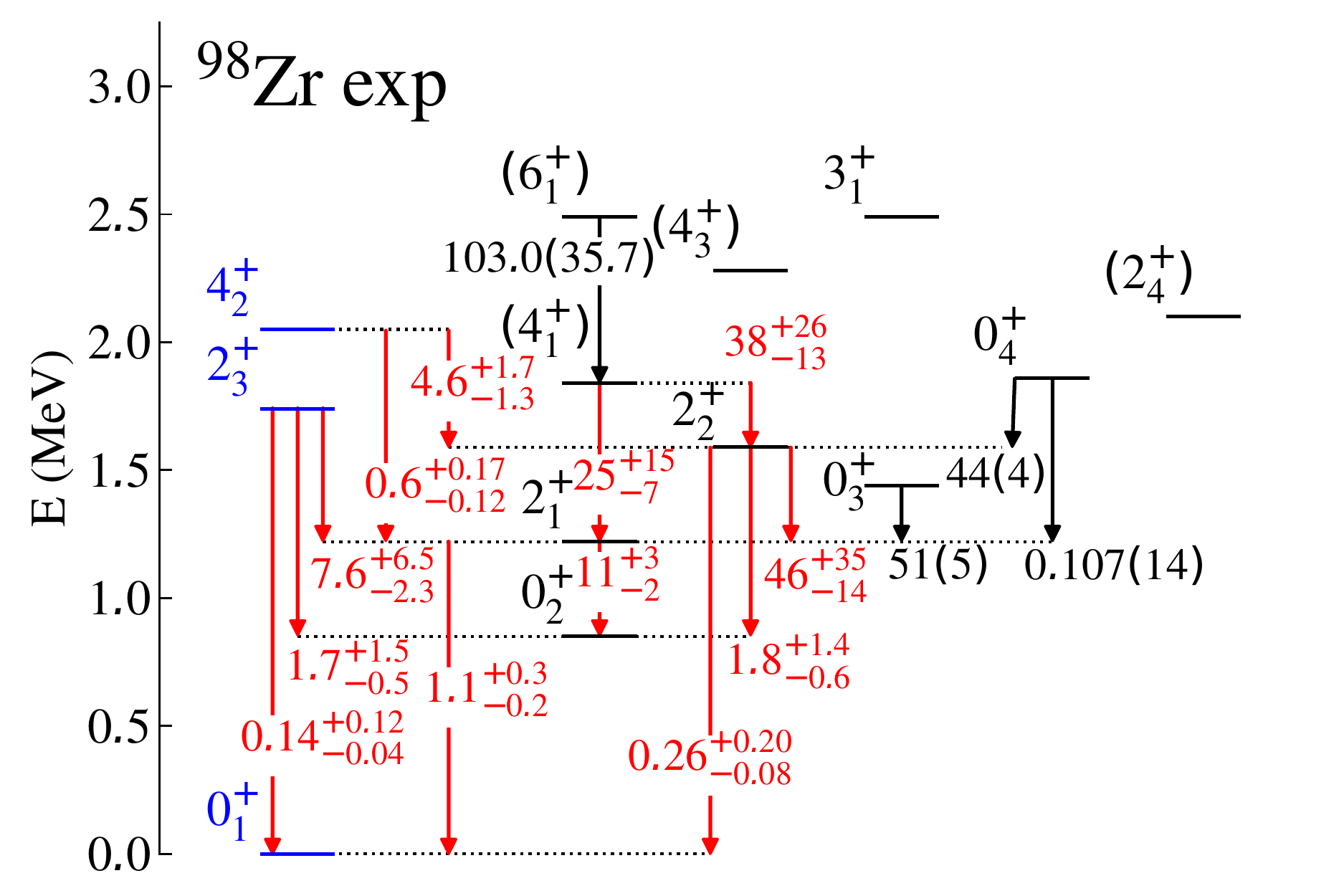}
\put (85,59) {(a)}
\end{overpic}
\begin{overpic}[width=0.49\linewidth]{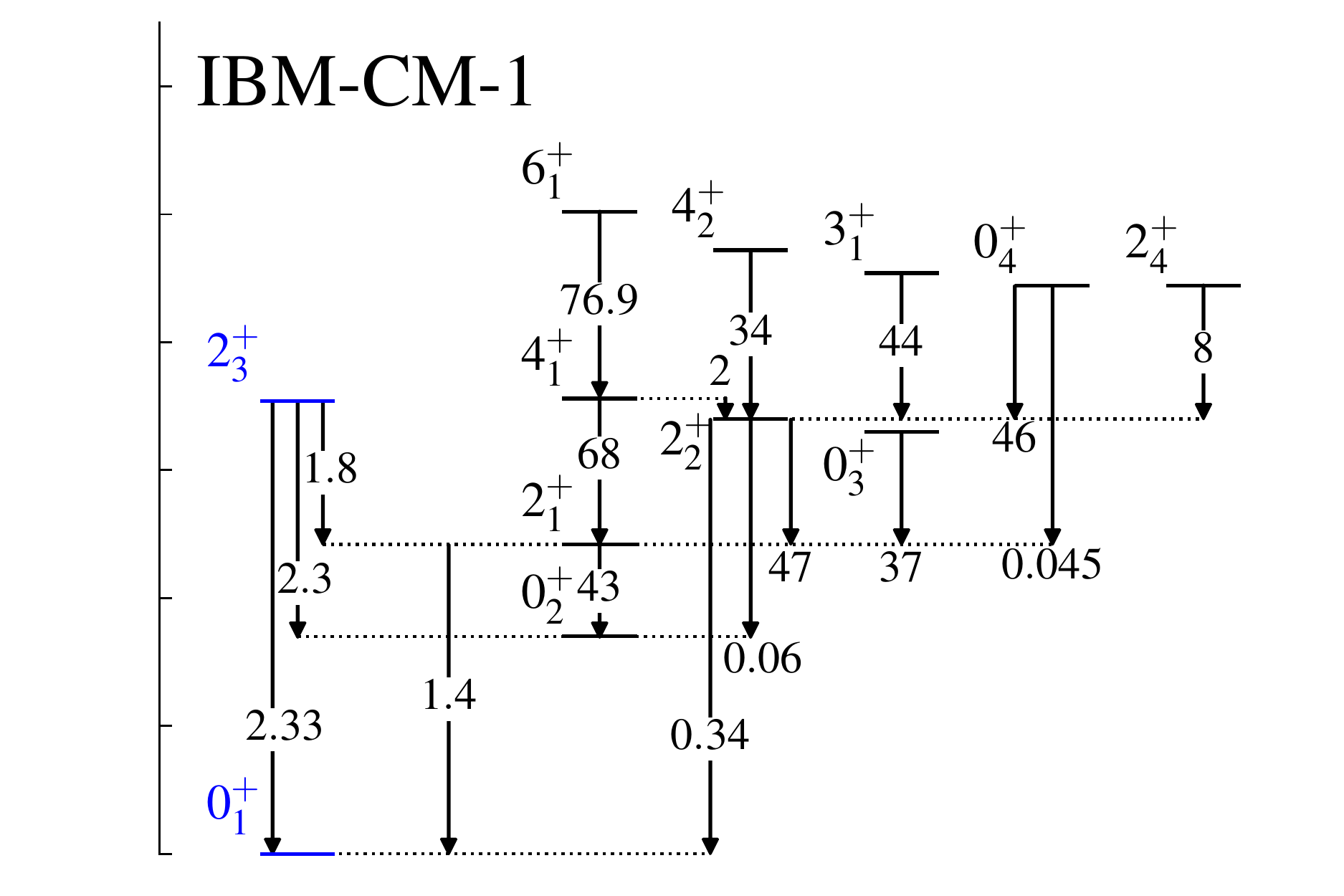}
\put (85,59) {(b)}
\end{overpic}
\caption{\label{fig:spectrum_th}
\small
(a) Experimental and (b) calculated \cite{Gavrielov2019, Gavrielov2020} energy levels in MeV and $E2$ rates in W.u. for $^{98}$Zr. Levels marked in blue (black) indicate states assigned to the $A$-normal ($B$-intruder) configuration. $E2$ transitions strengths marked in red are from the current work.}
\end{figure*}

In Fig. \ref{fig:spectrum_th}, we compare the {IBM-CM} calculation of \cite{Gavrielov2019, Gavrielov2020} (named {IBM-CM-1}) to the present new experimental results for $^{98}$Zr. The spectrum is divided into sectors of normal states (in blue) and intruder states (in black).
The $0^+_1$ and $2^+_3$ states are calculated in the {IBM-CM-1} to be normal states, part of a seniority-like spectrum of neutron single-particle excitations, which is mostly outside the IBM model space. Therefore, the experimental $4^+_2$ level is not considered in the calculation. The remaining states, shown in Fig.~\ref{fig:spectrum_th}, have an intruder character and are calculated to be quasi-spherical or weakly-deformed. Accordingly, the experimental $0^+_2,~2^+_1,~(0^+_3,~2^+_2,~4^+_1),~(0^+_4,~2^+_4,~3^+_1,~4^+_3,~6^+_1)$ states correspond to calculated states dominated by U(5) components with $n_d\approx0,1,2,3$, respectively, within the intruder part of the wave function $\ket{\Psi_B;[N_b+2],L}$. The resulting mixing between the two configurations is weak, \textit{e.g.}, $a^2\!=\!98.2\%$ for the ground state ($0^+_1$) and $b^2\!=\!98.2\%$  for the intruder state ($0^+_2$).
These findings result in an agreement with the new experimental results of the current work, as seen in Table~\ref{t:be2} and Fig.~\ref{fig:spectrum_th}. The weak $E2$ rates $\be{2}{2}{0}{2}\!=\!1.8^{+1.4}_{-0.6}$~W.u. and strong $E2$ rates $\be{2}{2}{2}{1}\!=\!46^{+35}_{-14}$~W.u. conform with the {IBM-CM-1} interpretation of quasi-phonon structure for the intruder band. This interpretation agrees also with the previously measured $E2$ rates for  $0^+_3\to2^+_1,~0^+_4\to2^+_2,~0^+_4\to2^+_1$~\cite{NuDat_A98} and $6^+_1\to4^+_1$~\cite{Singh2018}, listed in Table~\ref{t:be2}.
The measured weak $E2$ rates $\be{2}{3}{0}{1}\!=\!0.14^{+0.12}_{-0.04}$, $\be{2}{3}{0}{2}\!=\!1.7^{+1.5}_{-0.5}$, $\be{2}{1}{0}{1}\!=\!1.1^{+3}_{-2}$~W.u. and $\be{2}{2}{0}{1}\!=\!0.26^{+0.20}_{-0.08}$~W.u.
conform with the IBM-CM-1 calculation~\cite{Gavrielov2019, Gavrielov2020} and the interpretation of the $0^+_1$ and $2^+_3$ as normal states with single-particle character, weakly mixed with the intruder states. The measured $E2$ rates $\be{2}{3}{2}{1}\!=\!7.6^{+6.5}_{-2.3}$ W.u. deviates from the calculated value of 1.8 W.u., however, a merely 1\% decrease in the parameter $\epsilon^{(A)}_d$ in the Hamiltonian \cite{Gavrielov2019} results in a calculated value of 6.1 W.u. for this transition, without affecting significantly the remaining transitions in Table~\ref{t:be2}.
As mentioned above, the experimental $4^+_2$ state is excluded from the {IBM-CM-1} model space, however, the transition rates involving it, $\be{4}{2}{2}{1}\!=\!0.6^{+0.17}_{-0.12}$~W.u. and $\be{4}{2}{2}{2}\!=\!4.6^{+1.7}_{-1.3}$~W.u., support the assignment of the experimental $4^+_2$ as a normal single-particle state, weakly mixed with the intruder band.

The IBM-CM-1 describes reasonably well most of the transitions listed in Table~\ref{t:be2}.
However, some of the newly measured transitions within the intruder-band, reported in the present work, exhibit marked differences from previous measurements and from both the {IBM-CM-1} and MCSM calculations.
Specifically, the value ${\be{2}{1}{0}{2}\!=\!11^{+3}_{-2}}$~W.u. is
significantly lower than the recently measured value of 28.3(60) W.u. in Ref.~\cite{Singh2018}, and conforms only with the lower (11.5 W.u.) and upper (71.3 W.u.) limits obtained in Refs.~\cite{Ansari2017} and \cite{Witt2018}, respectively. Furthermore, while both the calculated {IBM-CM-1} (43.39 W.u.) and MCSM (70 W.u.) values are in-between the experimental upper \cite{Witt2018} and lower \cite{Ansari2017} limits, yet they deviate considerably from the explicit values of \cite{Singh2018}, and of the current work.
These deviations are somewhat surprising, in view of the trend in the calculated $E2$ rates from the first $2^+$ state to the first $0^+$ state within the intruder $B$-configuration, as portrayed by dashed line in Fig.~\ref{fig:e2}. Since deformation is increased as neutrons are added \cite{Federman1979}, an increase of this $B(E2)$ value at neutron number 58 is expected \cite{Togashi2016, Gavrielov2019, Gavrielov2020}, when going from neutron number 56 to 60.

Additional discrepancies between calculated and measured values occur for transitions involving the $4^+_1$ state. Specifically, both measured transition rates $\be{4}{1}{2}{1}\!=\!25^{+15}_{-7}$~W.u. and $\be{4}{1}{2}{2}\!=\!38^{+26}_{-13}$~W.u. are strong, a situation that cannot be accommodated neither by the IBM-CM-1, nor by the MCSM calculations. In the {IBM-CM-1} calculation, these values are 68 and 2 W.u., respectively, and reflect the fact that both the $4^+_1$ and $2^+_2$ are members of the $n_d\approx2$ triplet of configuration~($B$) and are weakly mixed with states in the normal $A$-configuration. In such circumstances, these states cannot be connected by strong $E2$ transitions, which follow the selection rules $\Delta n_d=\pm1$.  As shown in Table~\ref{t:be2}, both versions of the MCSM calculations, MCSM-1 and MCSM-2, encounter a similar problem and cannot accommodate simultaneously two strong transitions from the $4^+_1$ state. 

Recently, another independent {IBM-CM} calculation (named {IBM-CM-2} in Table~\ref{t:be2}) was carried out by Garc\'ia-Ramos and Heyde \cite{Garcia-Ramos2019}. 
In the {IBM-CM-2} the structure of the $4^+_1$ is similar to that of {IBM-CM-1}, however the $2^+_1$ and $2^+_2$ states exhibit strong normal-intruder mixing with $a^2\!=\!55\%$ and $a^2\!=\!45\%$, respectively. Consequently, the {IBM-CM-2} can describe adequately the empirical $\be{4}{1}{2}{1}$ and $\be{4}{1}{2}{2}$ rates. However, this structure leads to other noticeable discrepancies. In particular, the calculated ${\be{2}{2}{0}{2}\!=\!47}$~W.u., ${\be{2}{3}{2}{1}\!=\!46}$~W.u. and ${\be{2}{2}{2}{1}\!=\!0.55}$~W.u. are at variance with the experimental values of $1.8^{+1.4}_{-0.6}$,  $7.6^{+6.5}_{-2.3}$ and $46^{+35}_{-14}$~W.u., respectively.

\definecolor{GreenNoam}{rgb}{0,0.5,0}
\begin{figure*}[h!]
\centering
\includegraphics[width=0.7\linewidth]{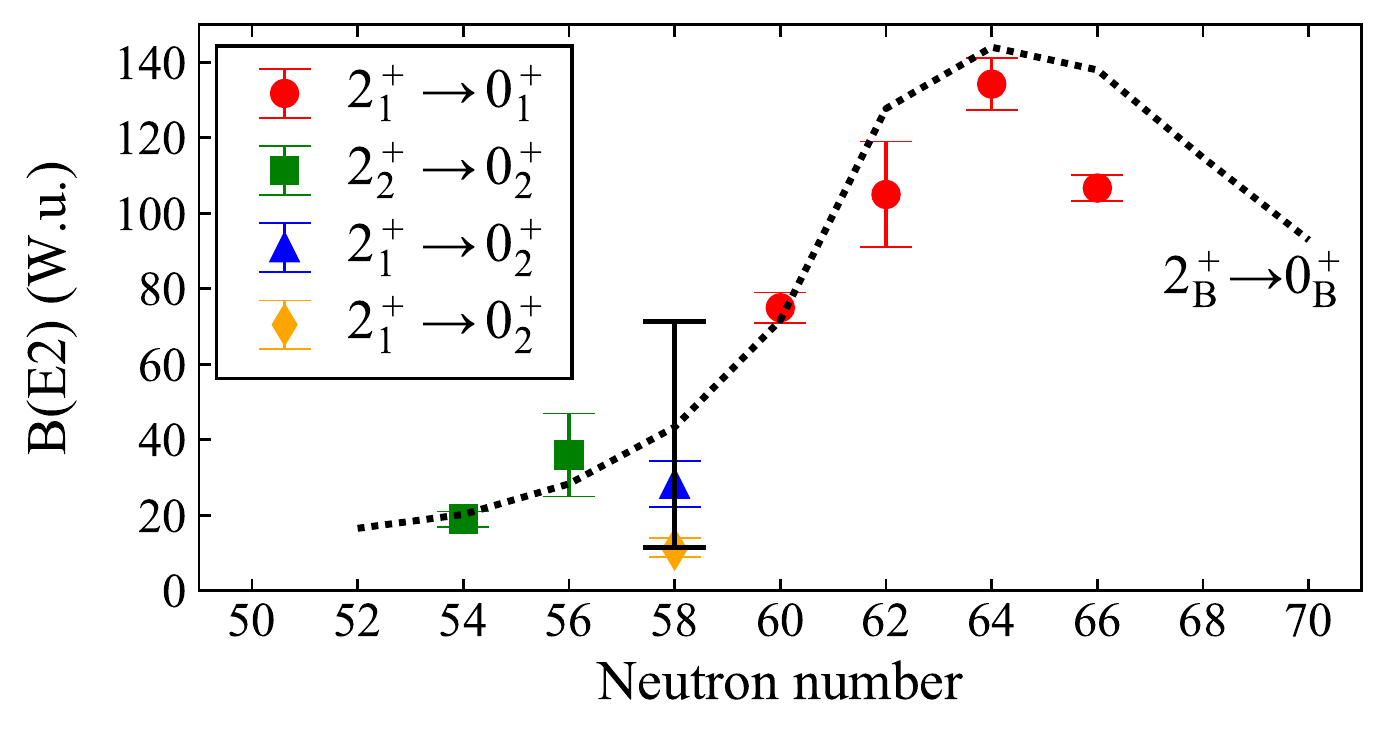}
\caption{\label{fig:e2}
\small
$E2$ transition rates in W.u. for $2^+\to 0^+$ transitions within the intruder $B$-configuration for the Zr isotopes. The symbols (${\color{red}\CIRCLE},~{\color{GreenNoam}\blacksquare},~{\color{blue}\blacktriangle},~{\color{orange}\blacklozenge}$) denote experimental rates. The dashed line depicts the IBM-CM-1 calculation of \cite{Gavrielov2019}.
The data for $^{94}$Zr, $^{96}$Zr, $^{100}$Zr, $^{102}$Zr and ($^{104}$Zr, $^{106}$Zr) are taken from \cite{Chakraborty2013,Kremer2016,Ansari2017,NuDat_A102,Browne2015}, respectively.
For $^{98}$Zr, the experimental values are from the current paper ({\color{orange}$\blacklozenge$}), from Ref. \cite{Singh2018} ({\color{blue}$\blacktriangle$}) and the black upper and lower limits are from Refs. \cite{Ansari2017,Witt2018}.
Note that the explicit experimental values in $^{98}$Zr deviate from one another as well as from the calculated values of the {IBM-CM-1} (43.39 W.u.) and of the MCSM (70 W.u.).}
\end{figure*}

\section{Conclusions and outlook}

The lifetimes of the $2^+_1$, $2^+_2$, $2^+_3$, $4^+_1$, $4^+_2$ and the $3^-_1$ states in $^{98}$Zr have been measured using the Doppler based techniques RDDS and DSA.
The results have been compared to the recently performed calculations in the framework of the Monte-Carlo shell-model and the interacting boson model with configuration mixing. Although both approaches provide a good overall description of the structure of $^{98}$Zr, there are some noticeable discrepancies. Some of the present measured transitions within the intruder-band exhibit marked differences from the previous measurements in Ref.~\cite{Singh2018}. 
Most notably, the measured weak transition $\be{2}{1}{0}{2}$ suggests the need to corroborate the lifetime of the $2^+_1$ state using other lifetime measurement methods and to explore further the structure of the $2^+_1$ and the $0^+_2$ states.  This suggests the need for further theoretical and experimental investigations. It would also be interesting to corroborate the $2^+_2 \rightarrow 0^+_3$ branching and the lifetime of the $2^+_2$ state to obtain the ${\be{2}{2}{0}{3}}$ value, which is calculated to be weak (6.54~W.u.) in the IBM-CM-1 and strong (49~W.u) in the MCSM. This can provide clues towards understanding whether the $0^+_3$,  $2^+_2$ and $4^+_1$ states are part of a quasi-two-phonon triplet as in the IBM-CM-1 calculation \cite{Gavrielov2019, Gavrielov2020} or part of a deformed configuration, possibly separated from the $0^+_2$ and the $2^+_1$ states as in the MCSM calculation \cite{Togashi2016, Singh2018}.

\section{Acknowledgments}

A.E. and V.K. acknowledge the financial support by the BMBF under Grant No. 05P19PKFNA. J.-M.R. and L.K. acknowledge the financial support by the Deutsche Forschungsgemeinschaft (DFG) under Grant No. JO391/16-2. We thank J.E. Garc\'ia-Ramos for providing the $B$($E2$) values from the calculation in Ref.~\cite{Garcia-Ramos2019}. We also thank T. Otsuka for providing the  $B$($E2$) values from the calculation in Ref.~\cite{Togashi2016}. The work of A.L. and N.G. is supported by the US-Israel Binational Science Foundation Grant No. 2016032 and in part by the Israel Science Foundation Grant No. 586/16.

\bibliography{./references}

\end{document}